\documentclass[a4paper]{article}

\usepackage[english]{babel}
\usepackage[utf8x]{inputenc}
\usepackage[T1]{fontenc}
\usepackage{multirow}
\usepackage{pifont}

\usepackage[a4paper,top=3cm,bottom=2cm,left=3cm,right=3cm,marginparwidth=1.75cm]{geometry}

\usepackage{amsmath}
\usepackage{graphicx}
\usepackage[colorinlistoftodos]{todonotes}
\usepackage[colorlinks=true, allcolors=blue]{hyperref}
\usepackage{authblk}
\usepackage{url}

\title{Microsoft Malware Classification Challenge}
\author[1]{{\normalsize Royi Ronen}}
\author[2]{{\normalsize Marian Radu\footnote{This work was done while this author was with Microsoft.}}}
\author[1]{{\normalsize Corina Feuerstein}}
\author[3]{{\normalsize Elad Yom-Tov}}
\author[4]{{\normalsize Mansour Ahmadi}}
\affil[1]{{\normalsize Microsoft}}
\affil[2]{{\normalsize CrowdStrike}}
\affil[3]{{\normalsize Microsoft Research}}
\affil[4]{{\normalsize Northeastern University}}

\begin{document}
\date{{\small \{royir,corinaf,eladyt\}@microsoft.com, marian.radu@crowdstrike.com, m.ahmadi@northeastern.edu}}
\maketitle

\begin{abstract}
The Microsoft Malware Classification Challenge was announced in 2015 along with a publication of a huge dataset of nearly 0.5 terabytes, consisting of disassembly and bytecode of more than 20K malware samples. Apart from serving in the Kaggle competition, the dataset has become a standard benchmark for research on modeling malware behaviour. To date, the dataset has been cited in more than 50 research papers. 
Here we provide a high-level comparison of the publications citing the dataset. The comparison simplifies finding potential research directions in this field and future performance evaluation of the dataset.
\end{abstract}

\section{Introduction}

In recent years, the malware industry has become a large and well-organized market \cite{shahi2009}. Well funded, multi-player syndicates heavily invest in technologies and capabilities built to evade traditional protection, requiring anti-malware vendors to develop counter-mechanisms for finding and deactivating them. In the meantime, they inflict significant financial loss to users of computer systems.

One of the major challenges that anti-malware software faces today are the vast amounts of data which needs to be evaluated for potential malicious intent. For example, Microsoft's real-time anti-malware detection products executes on over 600M computers worldwide \cite{msrt18}. This generates tens of millions of daily data points to be analyzed as potential malware. One of the main reasons for these high volumes of different files is that in order to evade detection, malware authors introduce polymorphism to the malicious components. This means that malicious files belonging to the same malware ``family'', with the same forms of malicious behavior, are constantly modified and/or obfuscated using various tactics, so that they appear to be many different files.

A first step in effectively analyzing and classifying such a large number of files is to  group them and identify their respective families. In addition, such grouping criteria may be applied to new files encountered on computers in order to detect them as malicious and of a certain family. To facilitate research in this area, especially in the development of effective techniques for grouping variants of malware files into their respective families, Microsoft provided the data science and security communities with a malware dataset of unprecedented size. Here we summarize the many uses of this dataset, published to date.

\section{Dataset}

The malware dataset is almost half a terabyte when uncompressed. It consists of a set of known malware files representing a mix of 9 different families. Each malware file has an identifier, a 20 character hash value uniquely identifying the file, and a class label, which is an integer representing one of the 9 family names to which the malware may belong (See Table~\ref{tab:training}). For each file, the raw data contains the hexadecimal representation of the file's binary content, without the header (to ensure sterility).  The dataset also includes a metadata manifest, which is a log containing various metadata information extracted from the binary, such as function calls, strings, etc. This was generated using the IDA disassembler tool. The original question posed to participants was to classify malware to one of the 9 classes. The dataset can be downloaded from the competition website.\footnote{https://www.kaggle.com/c/malware-classification/data} 

\begin{table}[tbp]
\centering
\caption{Malware families in the dataset}
\label{tab:training}
\begin{tabular}{lcc}
\noalign{\smallskip} \hline\noalign{\smallskip} 
Family Name    & \# Train Samples & Type \\
\hline\noalign{\smallskip} 
Ramnit         & 1541              &     Worm        \\
Lollipop       & 2478              &     Adware        \\
Kelihos\_ver3  & 2942              &     Backdoor        \\
Vundo          & 475               &     Trojan        \\
Simda          & 42                &     Backdoor        \\
Tracur         & 751               &     TrojanDownloader        \\
Kelihos\_ver1  & 398               &     Backdoor        \\
Obfuscator.ACY & 1228              &     Any kind of obfuscated malware       \\
Gatak          & 1013              &     Backdoor       \\
\hline\noalign{\smallskip} 
\end{tabular}
\end{table}

\section{Citations Comparison}

Since the end of the competition in April 2015, more than 50 research papers and thesis works cited the competition and the dataset. 
Among the citations, several papers are not in English, which we are unable to read \cite{russian2016,korean2016,french2015,NODE07303202}. 
The remaining articles can be divided into two principal classes. The first category of papers referenced the challenge to either perform an abstract comparison or highlight the importance of machine learning for malware classification in industry, where the size of data is huge \cite{7413680,DBLP:journals/corr/GarciaM16,imran2016,DBLP:journals/corr/WangGQ16,DBLP:journals/corr/GaoWQ17,8062202,ndss18,Scofield:2017:FML:3151137.3151142,10.1007/978-981-10-7080-8_17,DBLP:journals/corr/abs-1708-08042,webeye2017,Hansen2017,10.1007/978-3-319-60876-1_10,Oscar2017,Fields2016,Fowler2015,8102915,10.1007/978-3-319-73830-7_6}. 
Papers in the second category performed partial or complete evaluation on the dataset to verify the effectiveness and/or efficiency of their proposed approach for various tasks. We list the papers of the second category in Table~\ref{tab:comparison} sorted by the publication date. Moreover, we summarize the main contribution or focus of each paper to make higher level clusters. Feature engineering, feature selection/fusion, being scalable, being robust, malware authorship attribution, detecting concept drift, performing a measurement, similarity hashing, classification techniques and deep learning are the major contributions of the papers. The diversity of the contributions has made the dataset a benchmark for various tasks, helping researchers provide a standard for evaluation and comparison. 

\begin{table}[tbp]
\centering

\begin{tabular}{ccl|c}
\hline\noalign{\smallskip} 
Month & Year & Method & Focus/Contribution of Research \\
\noalign{\smallskip}\hline\noalign{\smallskip}
Mar & 2016 & Ahmadi et al. \cite{Ahmadi:2016:NFE:2857705.2857713} & Feature Engineering,Feature Fusion,Being Scalable \\
May & 2016 & Drew et al. \cite{7527757} & Feature Engineering,Being Scalable \\
Jul & 2016 & Hu et al. \cite{7523365} & Being Scalable \\
Jul & 2016 & Narayanan et al. \cite{7856826} & Feature Engineering \\
Jul & 2016 & Celik et al. \cite{DBLP:journals/corr/CelikMIPS16} & Being Robust  \\
Aug & 2016 & Zhang et al. \cite{7847046} & Being Scalable,Classification Techniques \\
Sep & 2016 & Bhattacharya et al. \cite{DBLP:journals/corr/BhattacharyaMBC16} &  Being Scalable,Classification Techniques \\
Oct & 2016 & Dinh et al. \cite{7723750} & Classification Techniques \\
Oct & 2016 & Wojnowicz et al. \cite{7796904} &  Feature Reduction \\
Nov & 2016 & Borbely \cite{Borbely2016} & Clustering Techniques \\
Dec & 2016 & Burnaev et al. \cite{7836677} & Classification Techniques  \\
Dec & 2016 & Alrabaee et al. \cite{10.1007/978-3-319-51966-1_17} & Malware Authorship Attribution \\
\noalign{\smallskip}\hline\noalign{\smallskip}
Jan & 2017 & Drew et al. \cite{Drew2017} & Being Scalable,Classification Techniques  \\
Jan & 2017 & Patri et al. \cite{Patri2017DiscoveringMW} & Classification Techniques  \\
Mar & 2017 & Hassen et al. \cite{Hassen:2017:SFC:3029806.3029824} & Feature Engineering,Being Scalable \\
Mar & 2017 & Celik et al. \cite{Celik:2017:FCP:3041008.3041018} & Being Robust  \\
May & 2017 & Yousefi-Azar et al. \cite{7966342} & Feature Engineering \\
Jun & 2017 & Kebede et al. \cite{8268747} & Deep Learning \\
Jul & 2017 & Yuxin et al. \cite{Yuxin2017} & Deep Learning \\
Aug & 2017 & Zhang et al. \cite{8029425} & Clustering Techniques \\
Aug & 2017 & Jordaney at al. \cite{203684} & Detecting Concept Drift  \\
Aug & 2017 & Raff et al. \cite{DBLP:journals/corr/abs-1708-03346} & Similarity Hashing \\
Oct & 2017 & Kim et al. \cite{Kim2017} & Deep Learning \\
Nov & 2017 & Rahul et al. \cite{10.1007/978-981-10-6898-0_19} & Deep Learning \\
Dec & 2017 & Bagga \cite{Bagga2017} & Measurement and Comparison  \\
Dec & 2017 & Gsponer et al. \cite{10.1007/978-3-319-71246-8_3} & Classification Techniques \\
Dec & 2017 & Hassen et al. \cite{Hassen2017} & Feature Engineering \\
Dec & 2017 & Fan et al. \cite{8064195} & Being Scalable \\
Dec & 2017 & Kim \cite{10.1007/978-981-10-7605-3_215} & Deep Learning  \\
\noalign{\smallskip}\hline\noalign{\smallskip}
Jan & 2018 & Hwang et al. \cite{fselection2018} & Feature Selection \\
Feb & 2018 & Yan et al. \cite{Yan2018scn} & Deep Learning \\
Feb & 2018 & Kreuk et al. \cite{2018arXiv180204528K} & Adverserial Examples,Deep Learning \\
Feb & 2018 & Hassen et al. \cite{2018arXiv180204365H}  & Open Set Recognition \\
\noalign{\smallskip}\hline
\noalign{\smallskip}
\end{tabular}

\smallskip
\caption{A comparison between research papers that have performed partial or complete evaluation on Microsoft malware classification challenge dataset.}
\label{tab:comparison}
\end{table}

\section{Conclusion and Future Directions}

In this paper, we provide a short description of the characteristics of the Microsoft Malware Classification Challenge dataset. This dataset is becoming a standard dataset with more than 50 papers citing it. We enumerated these references as much as possible and compared their main contributions with respect to the dataset. The comparison helps the understanding of what the existing contributions are, and what  the potential research directions can be.

The authors aim to keep the reference table updated. We encourage the community to cite this paper when using the dataset, and update us about such work so it can be added to this paper.

\bibliographystyle{plainurl}
\bibliography{sample}

\end{document}